\documentclass[
nofootinbib,
 amsmath,amssymb,
]{revtex4-1}

\usepackage[utf8]{inputenc}

\usepackage[margin=27mm]{geometry}
\usepackage{ragged2e}
\usepackage{graphicx}
\usepackage{subcaption}
\usepackage[english]{babel}
\usepackage{natbib}

\begin{document}


\title{Position-Dependent Diffusion induced Non-monotonic decay of Certain Non-Equilibrium Phenomena in Condensed Phase}

\author{Sagnik Ghosh}
\affiliation{Indian Institute of Science Education and Research, Pune-411008, India}
\email{sagnik.ghosh@students.iiserpune.ac.in}

\author{Alok Samanta \& Swapan K Ghosh  }
\affiliation{UM-DAE Centre for Excellence in Basic Sciences, University of Mumbai, Kalina, Santacruz (E), Mumbai-400098, India}%
\email{swapan.ghosh@cbs.ac.in}

\date{December 12, 2021}

\begin{abstract}
The dynamics of various optically controlled non-equilibrium phenomena in the condensed phase are studied using the Liouville equation. We study a projection of the same in a slow moving coordinate, identified as the Reaction Coordinate approach, with a position dependent diffusion coefficient. Introduction of position dependence is shown to induce non-monotonicity in relaxations of certain Non-equilibrium correlation functions, previously unexplored in the theoretical as well as experimental studies. This is in contrast to the exponential relaxation of its position independent analogue, irrespective of initial conditions. We characterize the dependence of this non-monotonicity on the strength of spatial inhomogeneity of diffusion and on the strength of the restoring forces and also indicate ranges of combinations where this feature is exhibited to pave the way for its experimental detection. 
\end{abstract}

\maketitle
\justifying

\section{Introduction}

The study of non-equilibrium processes in presence of condensed phases as a medium forms a frontier facet of condensed matter theory. The advancement of ultra-fast spectroscopy, such as the pump-probe method, where one excites a subsystem using a short pulse femto second laser, and observe the systems transient dynamics and subsequent relaxation to equilibrium, usually in terms of dynamics of its correlation functions. 

These experiments have led to explorations of optical control of various non-equilibrium processes in the condensed phase, such as solvation, symmetric and assymetric electron-transfers, two-step kinetics etc, especially in the context of their dependence on the excitation frequency. The early theoretical explorations in this field is due to Onsanger \cite{OnsangerI,OnsangerII}. The foundation of these treatments under Linear Response Theory (LRT) is based on the assumption that the nature of the fluctuations governing the non-equilibrium dynamics is essentially the same as of its equilibrium counterparts. Thus in its scope the dynamics remains essentially Markovian and fails to demarcate its dependence on the initial preparations, that are reported in various pump-probe type experiments in the context of solvation dynamics \cite{sen2006femtosecond, mondal2006ultrafast, mandal2008ultrafast, ghatak2011solvation, adhikari2007femtosecond, adhikari2009ultrafast} or in Electron Transfer \cite{fedunov2004effect, nicolet2005effect, vauthey2006investigations, mikhailova2004nonequilibrium, ivanov2003effect}. The Marcus Theory of Electron Transfer \cite{marcusI1956,MarcusII1957,marcusIII1957} also postulates that the attainment of critical configuration of the electrons is due to thermal fluctuations and hence is independent of the initial excitement. The Keldysh Field Theory approach to the same is also identical to Marcus Theory upto the first order of perturbation \cite{Hansen}. 

Newer Theories have been developed to address the same \cite{PatraSamantaGhosh2011}. \nocite{Zwanzig1961} Here one studies a projection of the general Liouville Dynamics to a slow moving co-ordinate of the phase space, namely the Reaction-Coordinate (RC), where the dynamical equation takes the form of a Heat equation in presence of a bound potential, thus encapsulating an interplay of the opposing effects of diffusion and restoring forces in the pseudo energy surface. Dynamics in this RC approach provides a quantitative justification of the related concept in theoretical chemistry, as well as is capable of resolving the issue of initial condition dependent relaxation of various non-equilibrium correlation functions \cite{SinjiniDi2019,Sayantani2020}. The effects of introducing reversibility \cite{DholeGuptaJPCLett,DholeJena2011} and anharmonicity \cite{Mitradip2020} has also been studied. However all these studies has employed the standard heat equation, with a constant uniform Diffusion Coefficient.

With the woke of better resolving techniques, there have been recent interest of exploring heterogenity in free diffusion with the incorporation of a position dependant diffusion constant \cite{wolfson2018comment, olivares2013direct, ljubetivc2014recovering, nagai2020position}. Here we investigate the same under the RC approach, under a simple Harmonic Bound Potential. We report that this model exhibits a previously unreported phase of solvation reactions, which is characterised by a non-monotonic decay of the non-equilibrium solvation correlation time, as well as of the higher moments. This phase is non-existent in homogeneous diffusion, in both the for bound and free cases and is thus uncaptured by the aforementioned theories as well as remained unexplored by the similar experiments. In the balance of this paper, we will briefly revisit the relevant theory and then study the parametric dependence of this non-monotonic phase on the strength of inhomogeneity and the restoring force. We show that, independent of the restoring force, inhomogeneity in diffusion is a necessary condition for the existence of this non-monotonic phase. We construct a phase diagram of this parametric dependence and indicate the ranges where this non-monotonic phase exists, in reduced units, with the fondest hope that it would pave the way for its experimental detection.

\section{Theory}

The non-equilibrium solvation processes as is captured by the aforementioned experiments are essentially a two step process. The femto-second pulse excites the the solute in presence of the condensed media from a ground state (A) to an solute+solvent condensed phase (A*). This process is known in the literature as the Frank-Cordon Excitation. The LASER source is then switched off and the system is relaxes towards equilibrium. Once the equilibrium is achieved the system is expected to remain there modulo small fluctuations around the same.

During the process of relaxation the excited state may undergo a de-excitation to ground state by emitting fluorescent radiation, which is the experimentally observed quantity in the spectroscopic experiments. The frequency of this radiation is related to the Non-Equilibrium Solvation time correlation function , $S(t)$ as follows,

\begin{equation}
    S(t) = \frac{\nu (t)- \nu (\infty)}{\nu (0)- \nu (\infty)} = \frac{\overline{\Delta E} (t)- \overline{\Delta E} (\infty)}{\overline{\Delta E} (0)- \overline{\Delta E} (\infty)} 
\end{equation}

$\overline{\Delta E}$ is further computed from the Phase space probability distribution $\rho (\Gamma,t|\Gamma_0,0)$,

\begin{equation}
    \overline{\Delta E}(t) =\int d\Gamma [V_e(Q)-V_g(Q)]\rho (\Gamma,t|\Gamma_0,0)
\end{equation}

Where $V_g,V_e$ respectively are the ground and excited state potential surfaces and $Q$ represents the set of nuclear coordinate in the phase space. The dynamics of the phase space density is governed by the Liouville Equation,

\begin{equation}\label{Liouville}
    \frac{\partial}{\partial t} \rho (\Gamma,t|\Gamma_0,0) = i \mathcal{L} \rho (\Gamma,t|\Gamma_0,0)
\end{equation}

where $\mathcal{L}$ is the excited state Liouville operator.

The ground state here is constructed based on energy conservation principle taking into account the excitation frequency $\lambda$,

\begin{equation}
    \rho (\Gamma,t|\Gamma_0,0)=\frac{\exp{(-\beta V_e(Q))}\delta(V_e(Q)-V_g(Q)-\frac{\hbar c}{\lambda})}{\int dQ \exp{(-\beta V_e(Q))}\delta(V_e(Q)-V_g(Q)-\frac{\hbar c}{\lambda})}
\end{equation}

Here our main focus is on the projection of this phase space dynamics on the one dimensional energy gap coordinate, $\epsilon = V_e(Q)-V_g(Q)$. This is obtained by constructing the following projection operator,.

\begin{equation}
    P(x) = \int d \epsilon \frac{(x*\delta(V_e(Q)-V_g(Q)-\epsilon))}{\int d \Gamma \exp{(-\beta H_g(\Gamma))} \delta(V_e(Q)-V_g(Q)-\epsilon)}
\end{equation}

Employing this projection on the Liuoville Equation gives rise to the Reaction Coordinate Approach \cite{PatraSamantaGhosh2011,Zwanzig1961}, and gives rise to the following effective dynamical equation with a general diffusion kernel $D(\epsilon,\epsilon'|t-t')$ and a bound potential $V(\epsilon)$.

\begin{widetext}
\begin{equation}\label{generaleq}
\begin{split}
\frac{\partial}{\partial t} P(\epsilon,t) =   \frac{\partial}{\partial \epsilon}\int_{-\infty}^{\infty}d\epsilon'\int_{0}^{t}dt' D(\epsilon,\epsilon'|t-t')\Big \{\frac{\partial}{\partial \epsilon'}+\frac{\partial}{\partial \epsilon''} V(\epsilon'')\Bigr|_{\substack{\epsilon''=\epsilon'}} \Big \}P(\epsilon',t') 
\end{split}
\end{equation}
\end{widetext}

Following the Markovian assumption of a space-time local Diffusion Kernel ammounts to the form,

\begin{equation}
    D(\epsilon,\epsilon'|t-t')=D(\epsilon)\delta(\epsilon-\epsilon')\delta(t-t')
\end{equation}

This converts the general integral-differential equation of (\ref{generaleq}), to the following PDE,

\begin{widetext}
\begin{equation}
\begin{split}\label{gen1}
\frac{\partial}{\partial t} P(\epsilon,t) =   \frac{\partial}{\partial \epsilon} D(\epsilon)\Big \{\frac{\partial}{\partial \epsilon}+\frac{\partial}{\partial \epsilon''} V(\epsilon'')\Bigr|_{\substack{\epsilon''=\epsilon'}} \Big \}P(x,t)
\end{split}
\end{equation}
\end{widetext}

Our first job is to solve for $P(\epsilon,t)$ numerically from this equation. We work with a harmonic bound potential $V(\epsilon)=\frac{k}{2}\epsilon^2$, characterised by the restoring force strength  $k$ and the position dependent diffusion coefficient,

\begin{equation}
    D(\epsilon)=D_0(1+\alpha \epsilon^2)
\end{equation}

In the following we have studied the numerical solution to \ref{gen1} and dynamics of the Non-equilibrium solvation time correlation function $S(t)$ defined as follows,

\begin{equation}
    S(t)=\frac{1}{\epsilon_0}\int_{-\infty}^{\infty} d\epsilon\; \epsilon P(\epsilon,t|\epsilon_0,t_0)
\end{equation}

where $\epsilon_0$ is the energy gap related to the state of initial excitation. We exhibit in the following sections that dynamics governed by the (\ref{gen1}) is capable of exhibiting non-monotonic relaxation of $S(t)$ for a wide range of parameters $\alpha, k$ given a finite $D_0$, which arises exclusively from the inhomogenity of the diffusion constant.

\section{Non-monotonic Relaxation}

We start with an initial Gaussian Distribution, centered at $\epsilon_0$, work with vanishing boundary conditions at large $|\epsilon|$ and treat the Dirac-$\delta$ contribution of source as a limit of Gaussian, with vanishing variance.

\begin{equation}
    P(\epsilon_0,0)=\frac{1}{\sqrt{\pi}\sigma} \exp{\Big(\frac{\epsilon-\epsilon_0}{\sigma}\Big)^2}
\end{equation}

\begin{equation}
\begin{split}
    P(\infty,t)=0\\
    P(-\infty,t)=0\\
\end{split}
\end{equation}

For the following studies we have used a $\sigma=0.1$ for the initial distribution, and for all practical purposes truncated spatial infinities at $\epsilon=\pm 10$ in reduced units

\begin{equation}
\begin{split}\label{boundarycdn}
    P(10,t)=0\\
    P(-10,t)=0\\
\end{split}
\end{equation}

We numerically solve the PDE using a dynamic grid in Mathematica with maximal grid spacing $h=0.01$ in both spatio-temporal directions. Equation (\ref{gen1}) describes time evolution of states in various non-equilibrium processed in condensed phases such as solvation. In the case of solvation the physically interesting quantity which characterises the out of equilibrium effects is the Non-Equilibrium Solvation time correlation function, 

\begin{equation}
    S(t)=\frac{1}{\epsilon_0}\int_{-\infty}^{\infty} d\epsilon\; \epsilon P(\epsilon,t|\epsilon_0,t_0)
\end{equation}

\begin{widetext}

\begin{figure}[t]
\begin{subfigure}{0.32\textwidth}
\includegraphics[width=\linewidth]{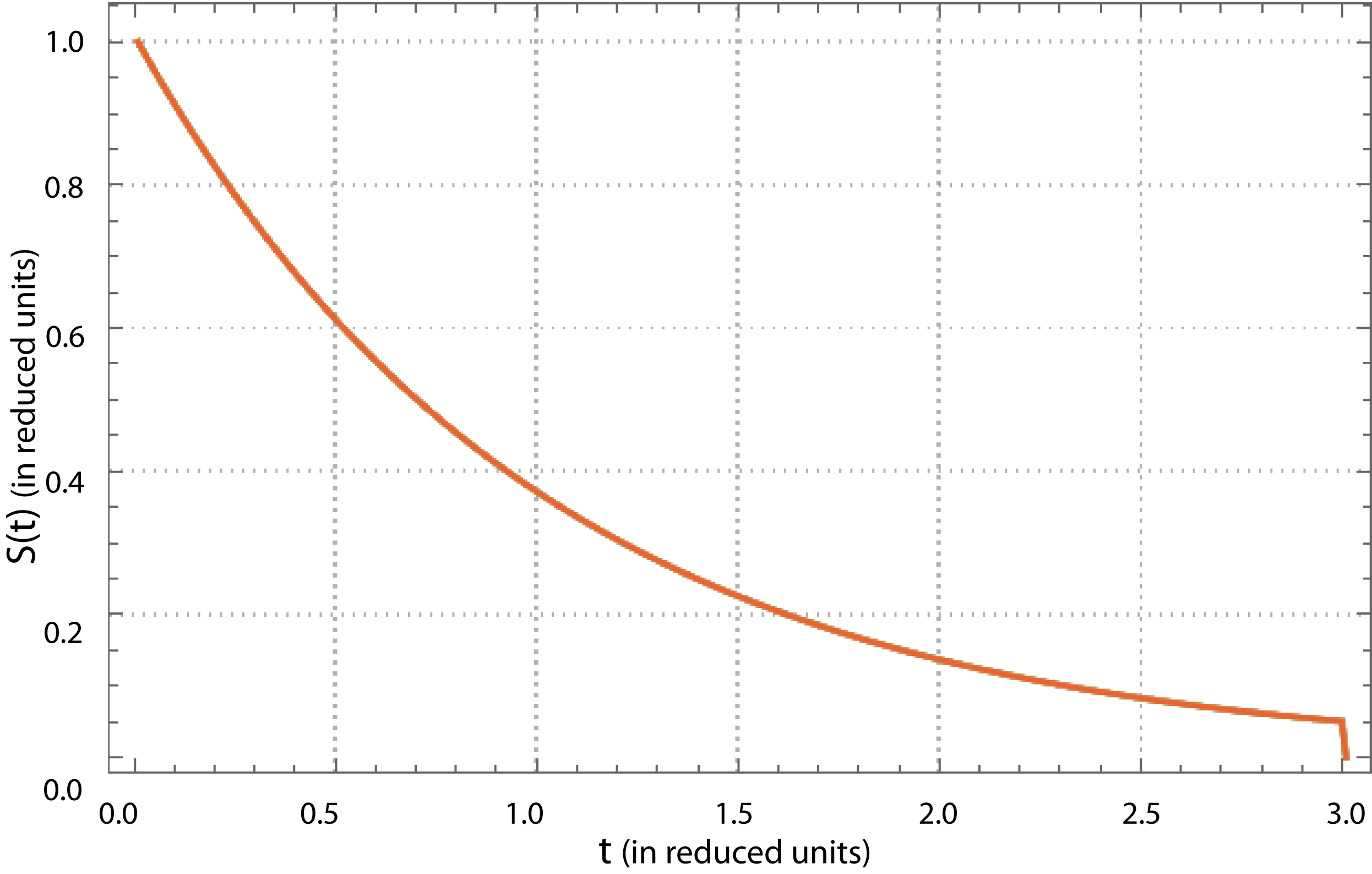} 
\caption{$\alpha=0, k=1$}
\label{fig:subim1a}
\end{subfigure}
\begin{subfigure}{0.32\textwidth}
\includegraphics[width=\linewidth]{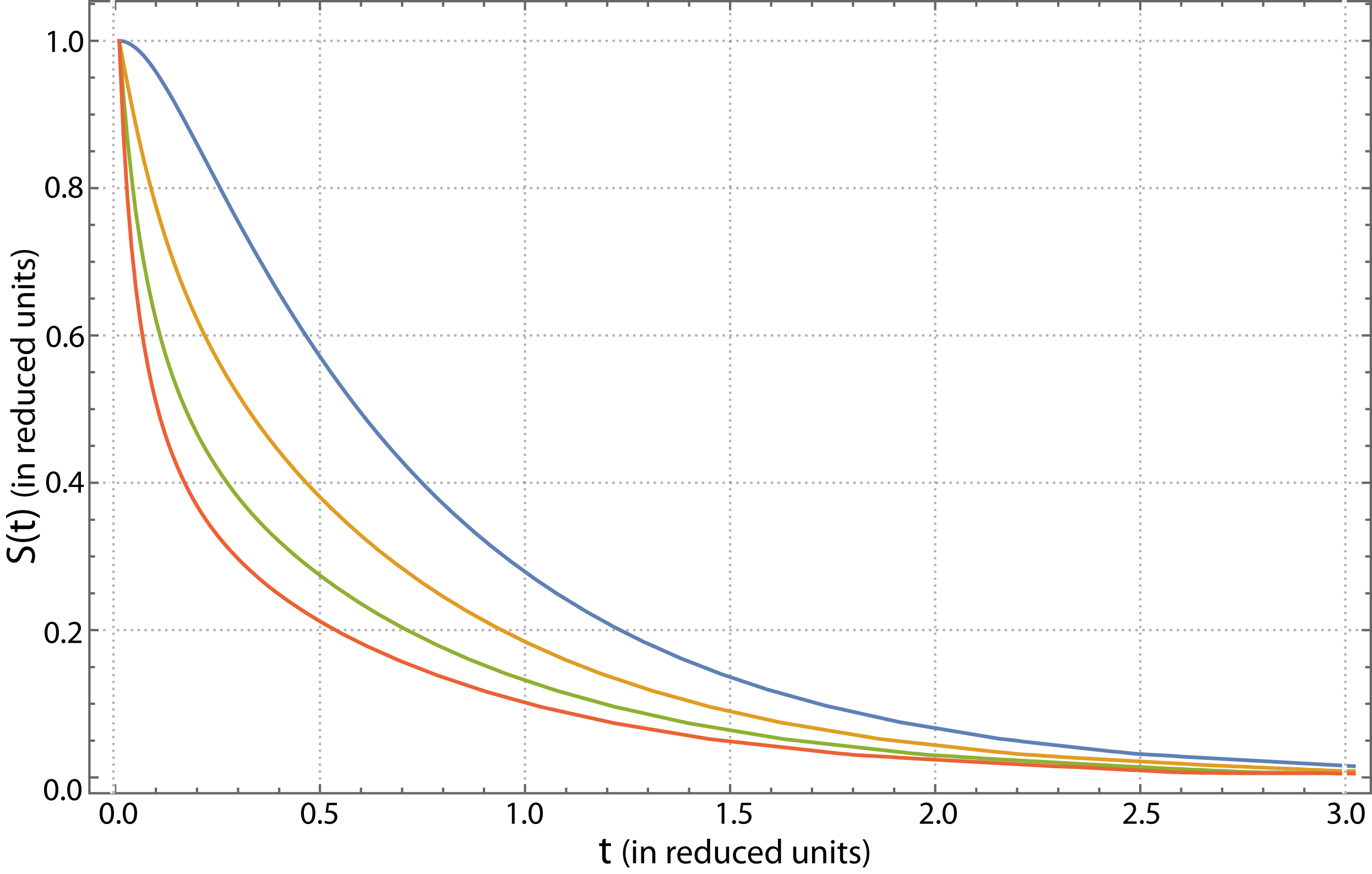} 
\caption{$\alpha=0, k=1$}
\label{fig:subim1b}
\end{subfigure}
\begin{subfigure}{0.32\textwidth}
\includegraphics[width=\linewidth]{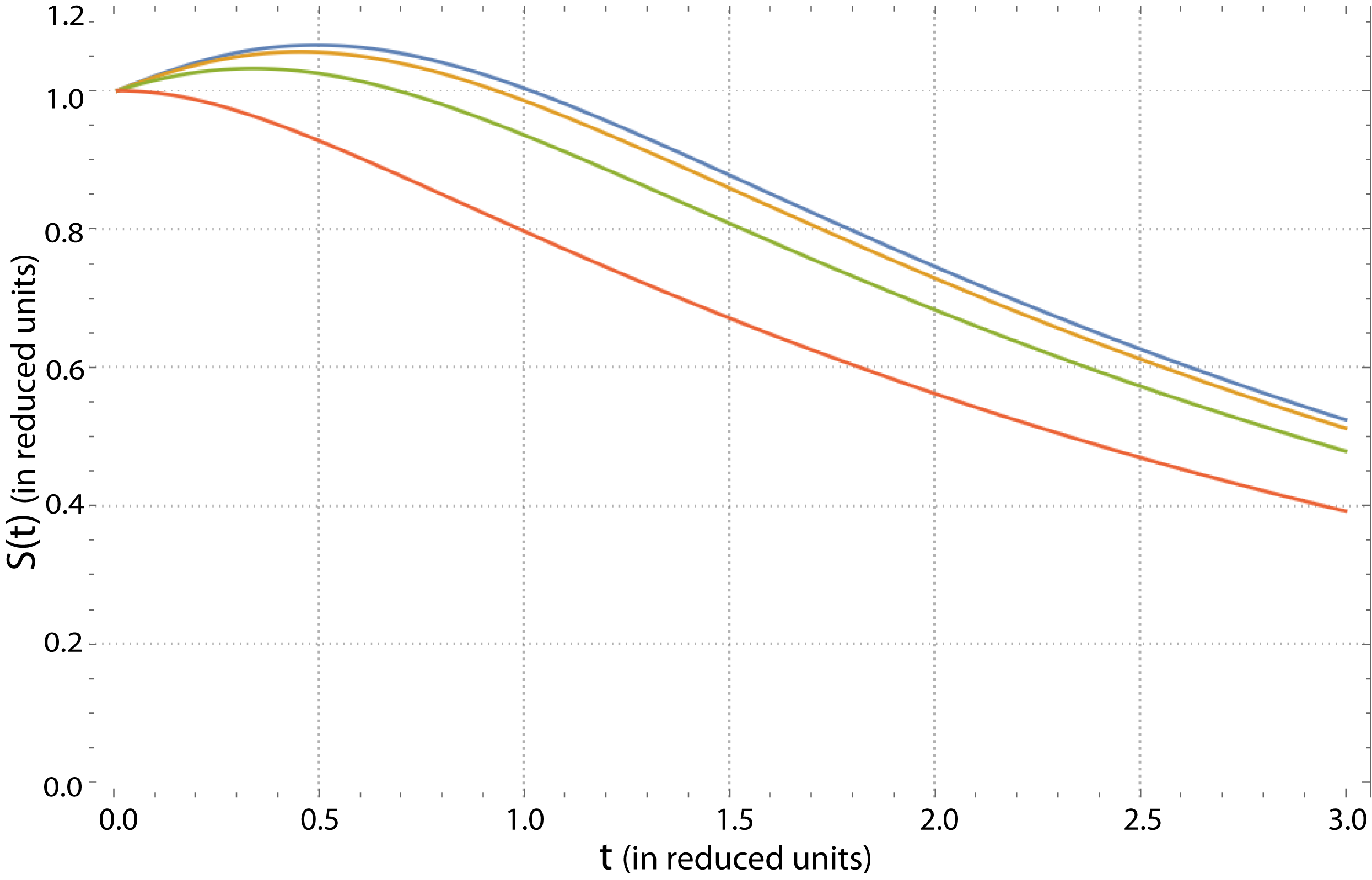}
\caption{$\alpha=0.25, k=0.25$}
\label{fig:subim1c}
\end{subfigure}

\caption{Comparison plot of the Non-equilibrium solvation correlation time $S(t)$ vs $t$ for various initial energy $\epsilon^0$, for the indicated combinations of $(\alpha,k)$. \textbf{Figure 1a.} shows the limiting case of position independent Diffusion Constant in a Harmonic Potential. From Linear Response Theory the relaxation of $S(t)$ to the equilibrium value in this case is known to be exponential and independent of the initial energy level ($\epsilon^0$). \textbf{Figure 1b.} exhibits that introduction of position dependence in diffusion constant breaks this universality and the relaxation is faster for higher initial energy levels. \textbf{Figure 1c.} reports the same for an intermediate combination $(\alpha=0.25,k=0.25)$. This case however depicts a \textit{non-monotonic decay}, where $S(t)$ first increases, attains a maximum value $S_{Max}$ and then relaxes to the equilibrium value, for a large range of initial energy levels. We identify it as a new phase and delve into characterizing the same in the balance of this paper.}
\label{fig:image1}

\end{figure}
\end{widetext}

At large time it is expected to relax at its equilibrium value. Figure (\ref{fig:subim1a}) reports dynamics of $S(t)$ for various initial conditions ($\epsilon^0$) for three combinations of ($\alpha,k$). The first case with $\alpha=0$ corresponds to the case of position independent diffusion. In this case the result is well known, and $S(t)$ exhibits an exponential decay, with a relaxation coefficient independent of the position of the initial wavelet. 

Introduction of spatial in-homogeneity lifts this degeneracy (Figure \ref{fig:subim1b}). With $\alpha=1, k=1$, the correlation time corresponding to wavelets initiated at $\epsilon^0=1,2,3,4$ reduced units respectively exhibits exponential relaxations with distinct rates. Further away from the ground state the system is prepared, faster is its relaxation to equilibrium value. However, at the intermediate range, $\alpha=0.25, k=0.25$ (Figure \ref{fig:subim1b}), for some initial conditions closer to the ground state we see a non-monotonic relaxation of the Non-equilibrium relaxation time.

\begin{widetext}

\begin{figure}[b]
\begin{subfigure}{0.32\textwidth}
\includegraphics[width=\linewidth]{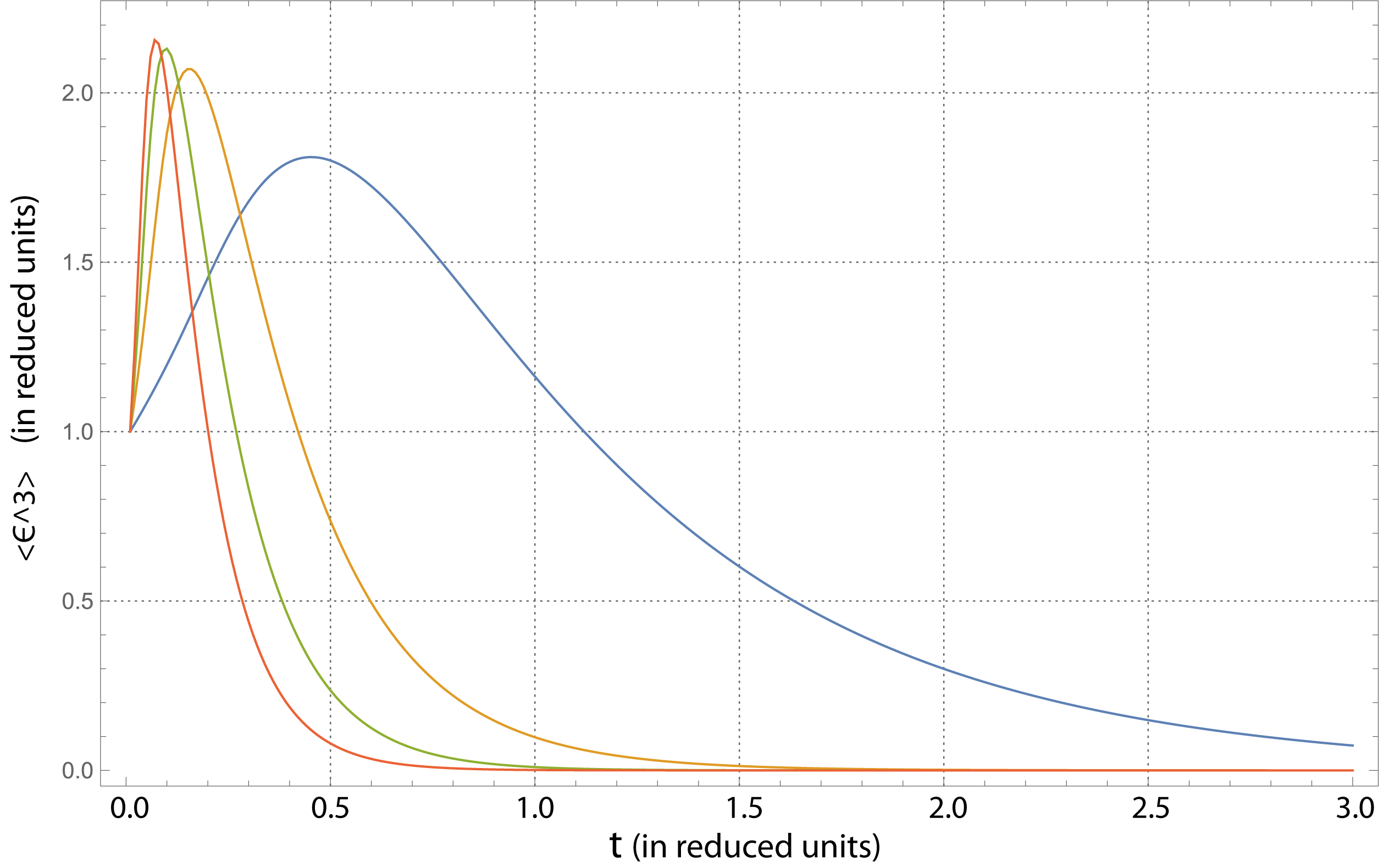} 
\caption{$<\epsilon^3>$}
\label{fig:subim2a}
\end{subfigure}
\begin{subfigure}{0.32\textwidth}
\includegraphics[width=\linewidth]{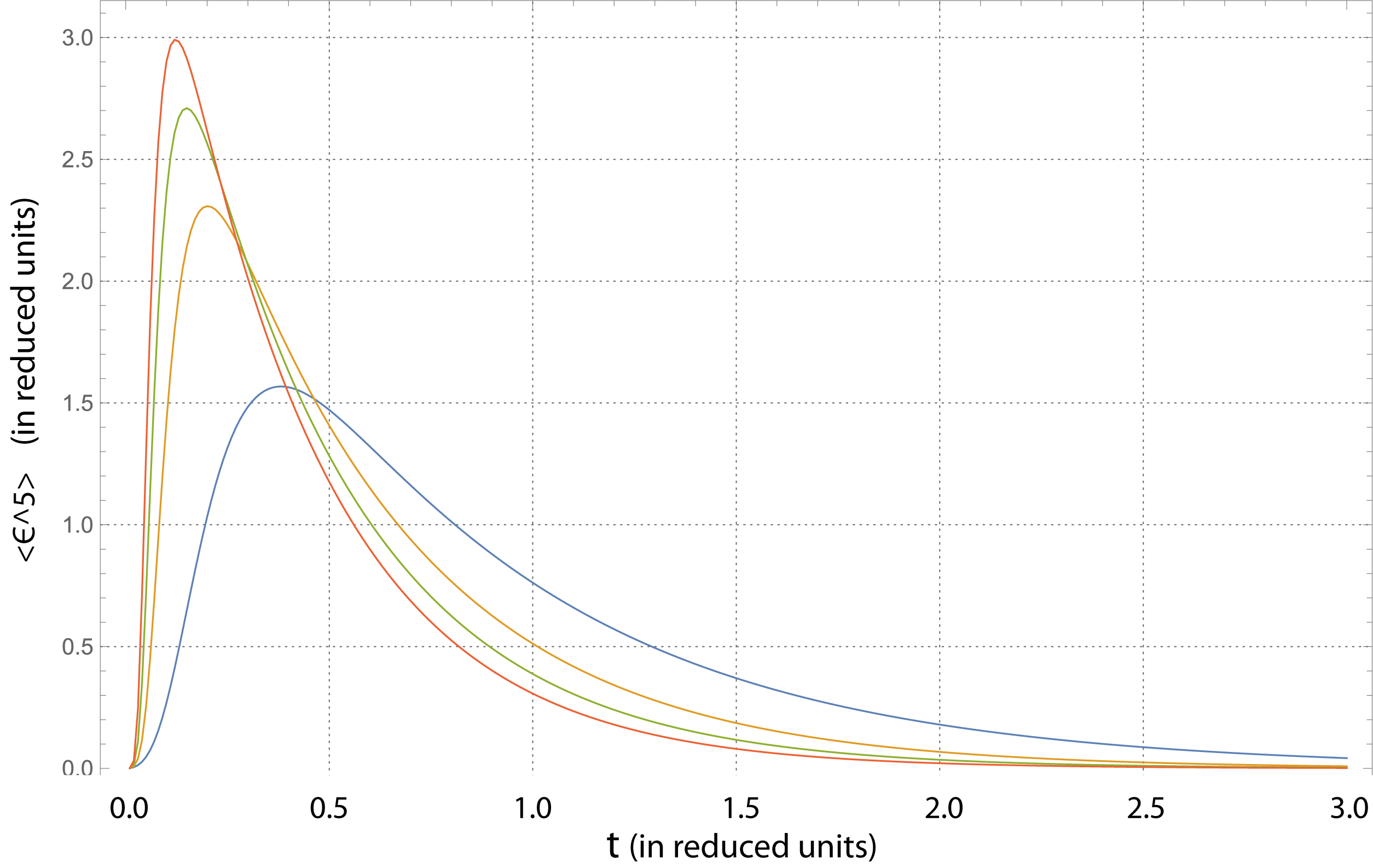} 
\caption{$<\epsilon^5>$}
\label{fig:subim2b}
\end{subfigure}
\begin{subfigure}{0.32\textwidth}
\includegraphics[width=\linewidth]{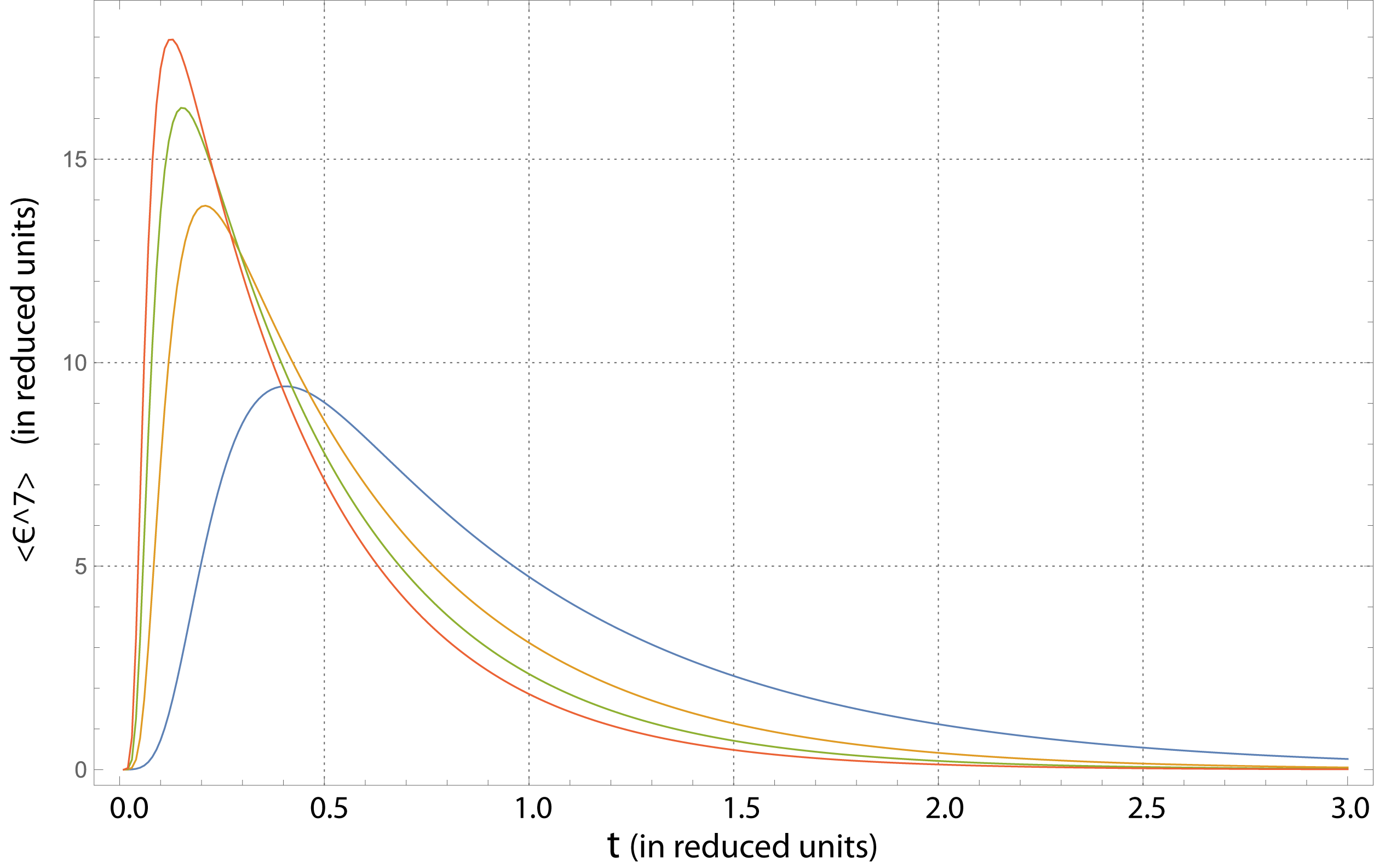} 
\caption{$<\epsilon^7>$}
\label{fig:subim2c}
\end{subfigure}
\begin{subfigure}{0.32\textwidth}
\includegraphics[width=\linewidth]{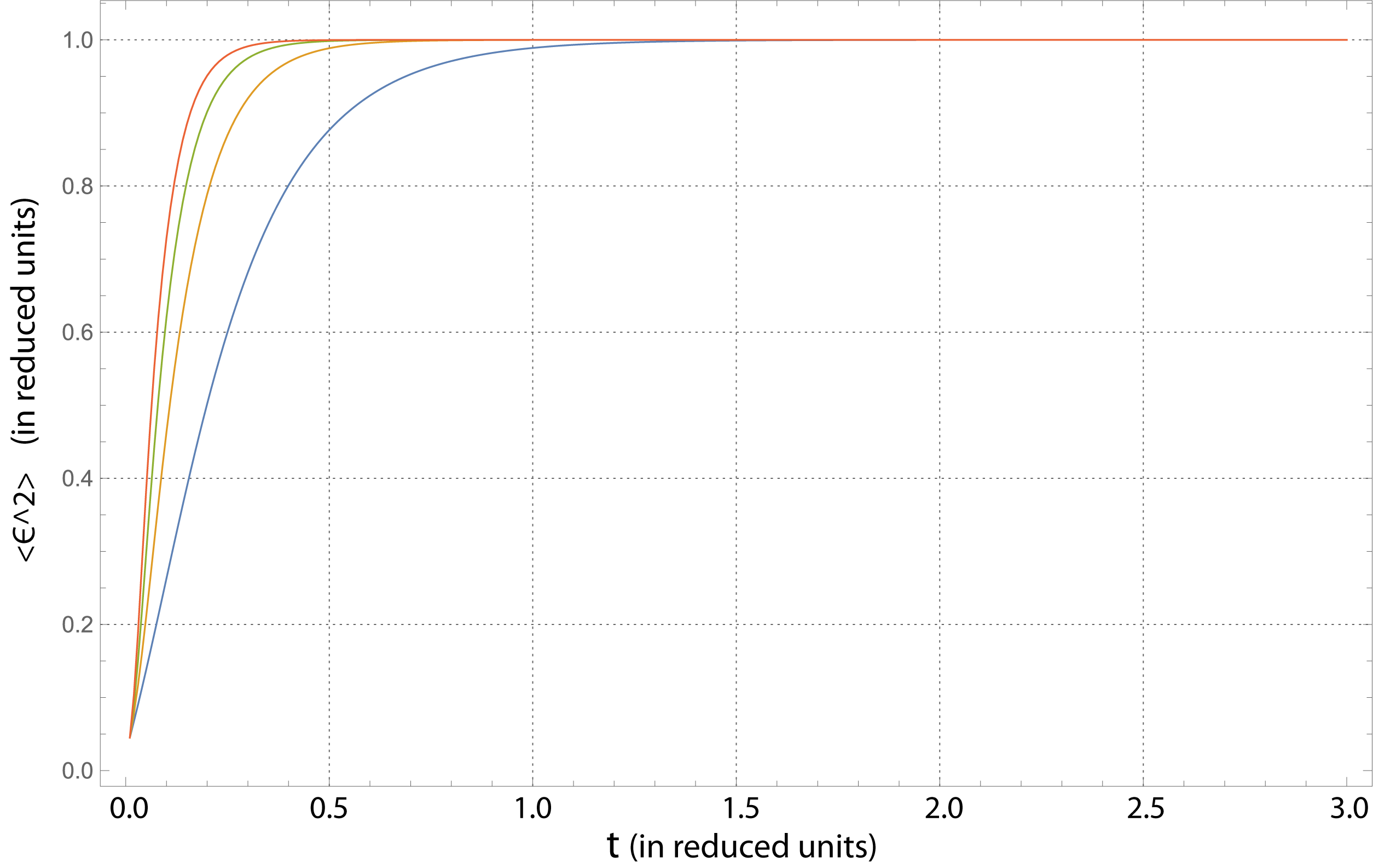} 
\caption{$<\epsilon^2>$}
\label{fig:subim2d}
\end{subfigure}
\begin{subfigure}{0.32\textwidth}
\includegraphics[width=\linewidth]{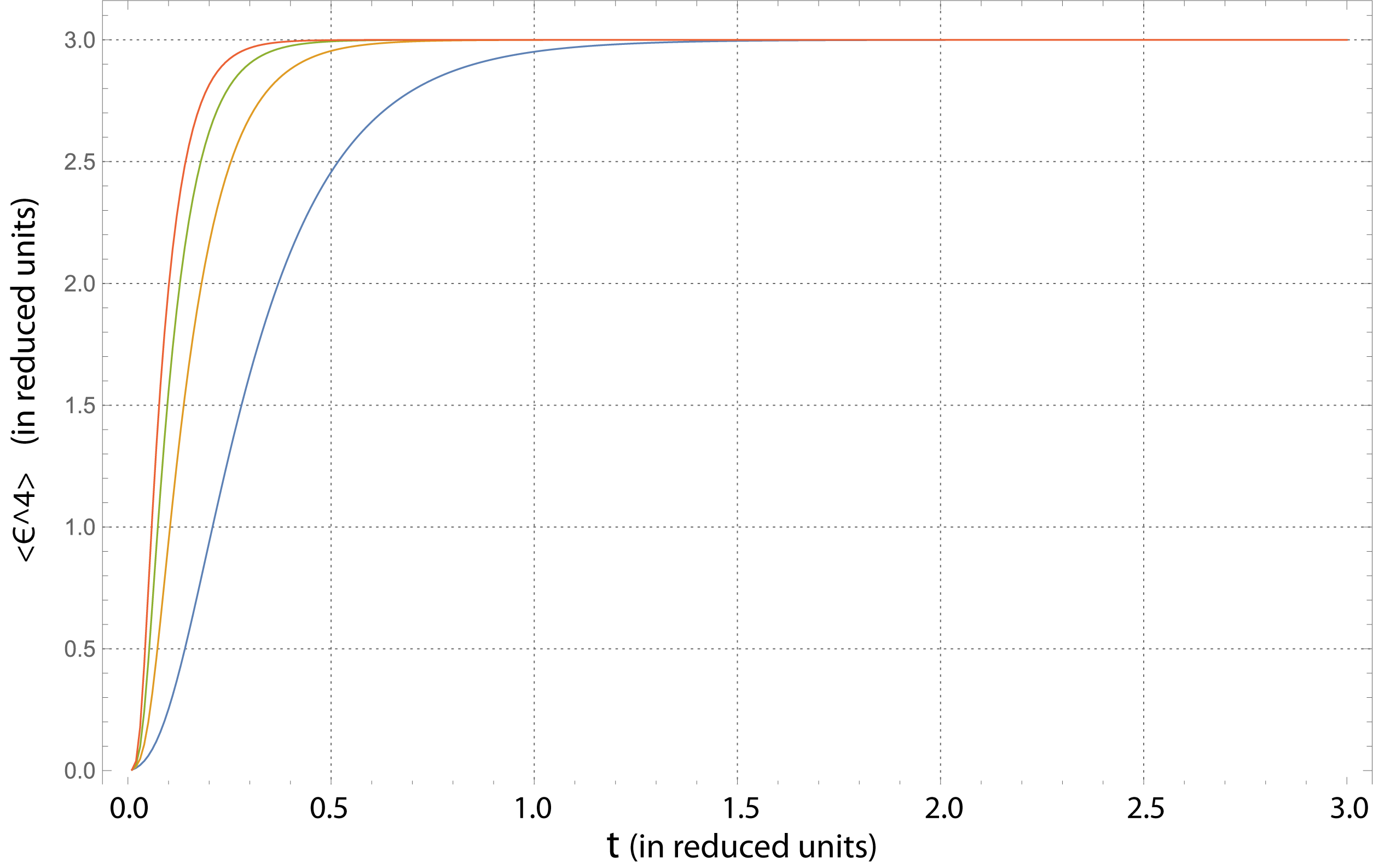} 
\caption{$<\epsilon^4>$}
\label{fig:subim2e}
\end{subfigure}
\begin{subfigure}{0.32\textwidth}
\includegraphics[width=\linewidth]{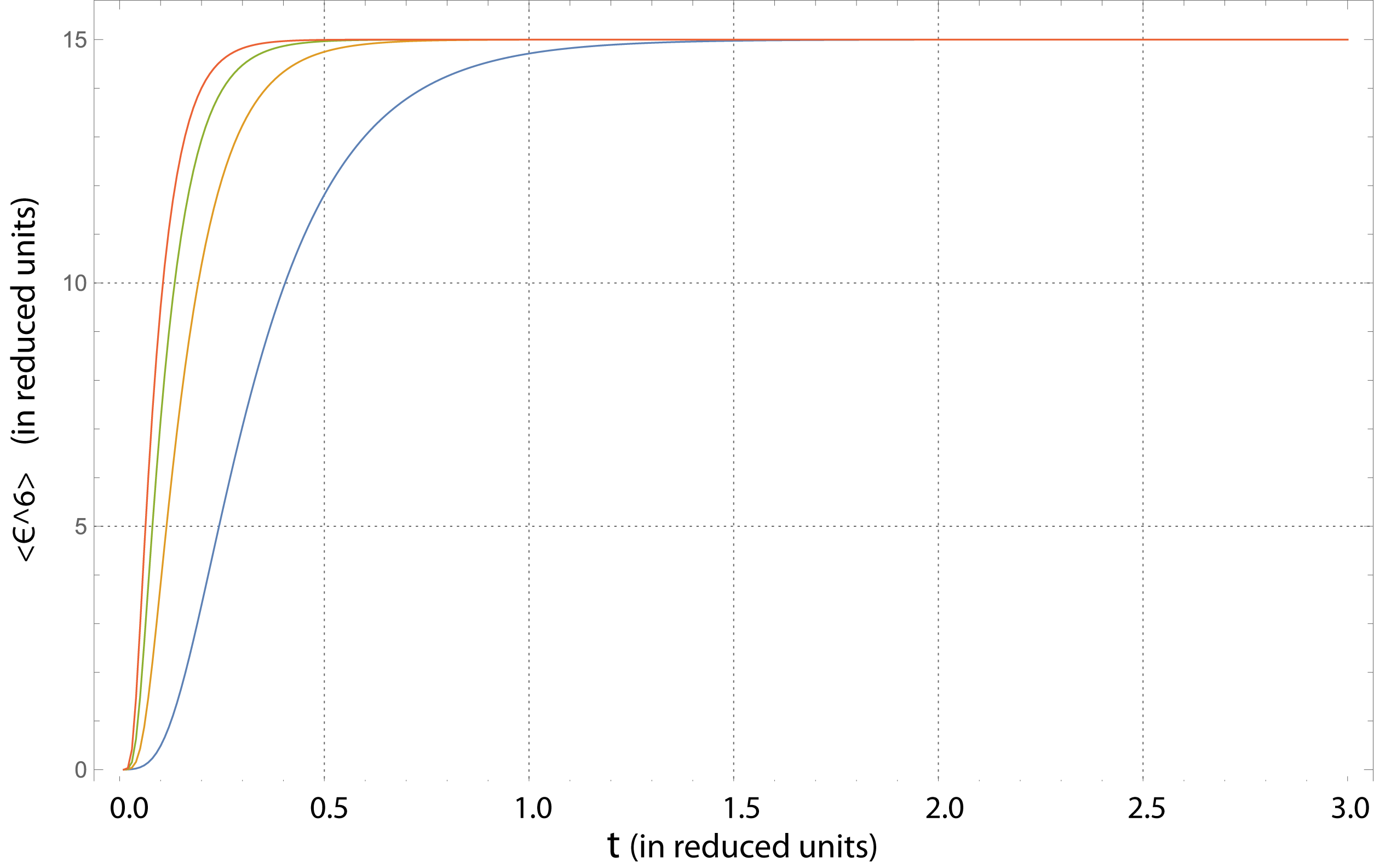} 
\caption{$<\epsilon^6>$}
\label{fig:subim2f}
\end{subfigure}

\caption{Comparison of various higher moments of the energy gap vs t for different alpha (1,4,7,10 blue to orange) For various initial wavelet as indicated K 1}
\label{fig:image2}

\end{figure}
\end{widetext}

This effect forms the thesis of our work presented here. As as is depicted later in the study, this phase is prevalent for a wide range of initial conditions, spatial inhomoeneity of the Diffusion Coefficient as well as the restoring strengths. This effect is also robust in the higher moments of the energy difference.  
In Figure (\ref{fig:image2}), we compute the next six moments for $\epsilon^0=0.2 k=1$ and for various $\alpha$ (1,4,7,10), where the first moment is known to exhibit the non-monotonic relaxation. It is shown that all the odd moments also exhibit non-monotonic relaxations for the same parameters, whereas the even moments exhibit an exponential relaxtion. 

\begin{widetext}

\begin{figure}[t]
\begin{subfigure}{0.49\textwidth}
\includegraphics[width=\linewidth]{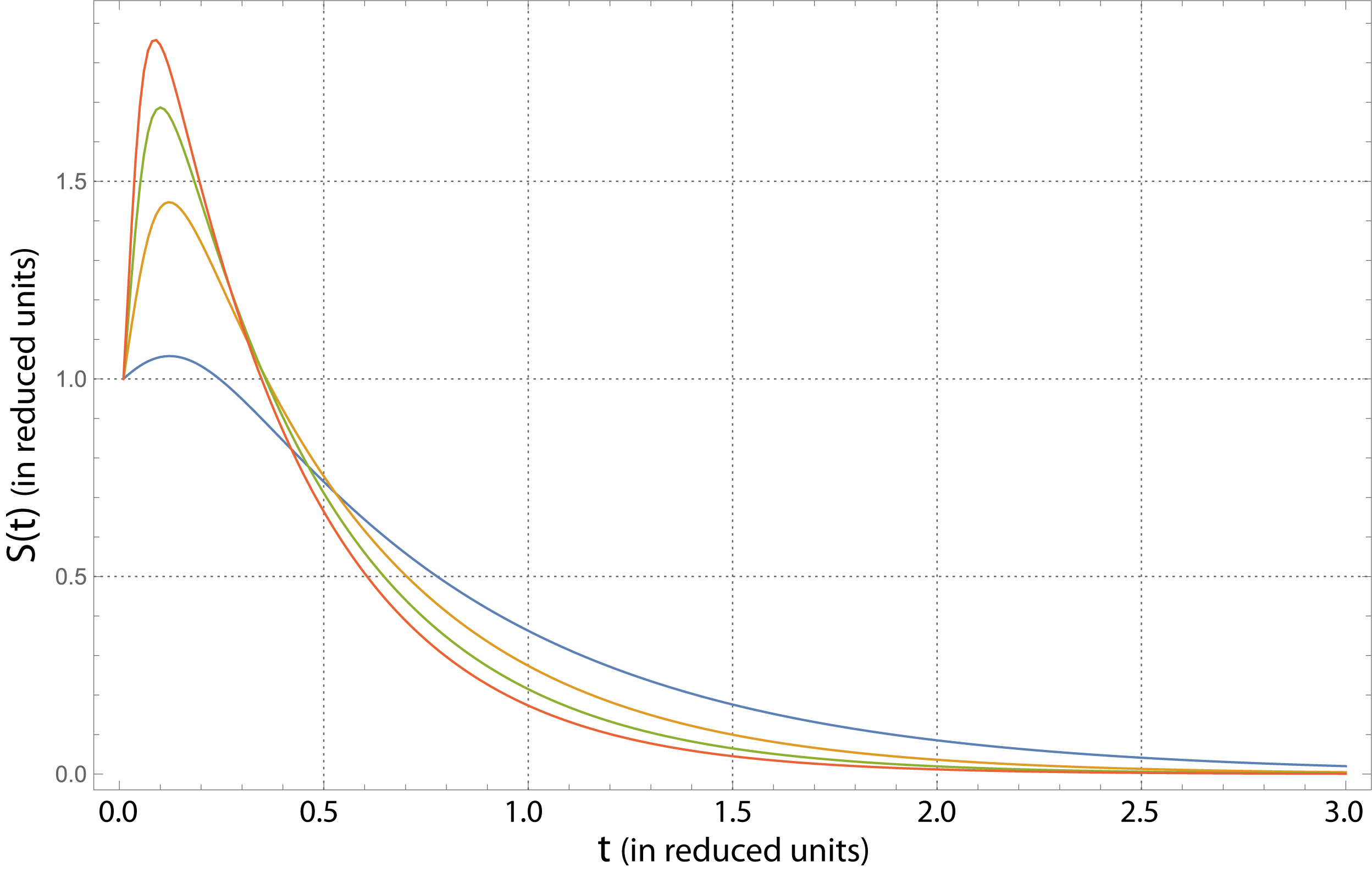} 
\caption{$\epsilon^0=0.2$}
\label{fig:subim3a}
\end{subfigure}
\begin{subfigure}{0.49\textwidth}
\includegraphics[width=\linewidth]{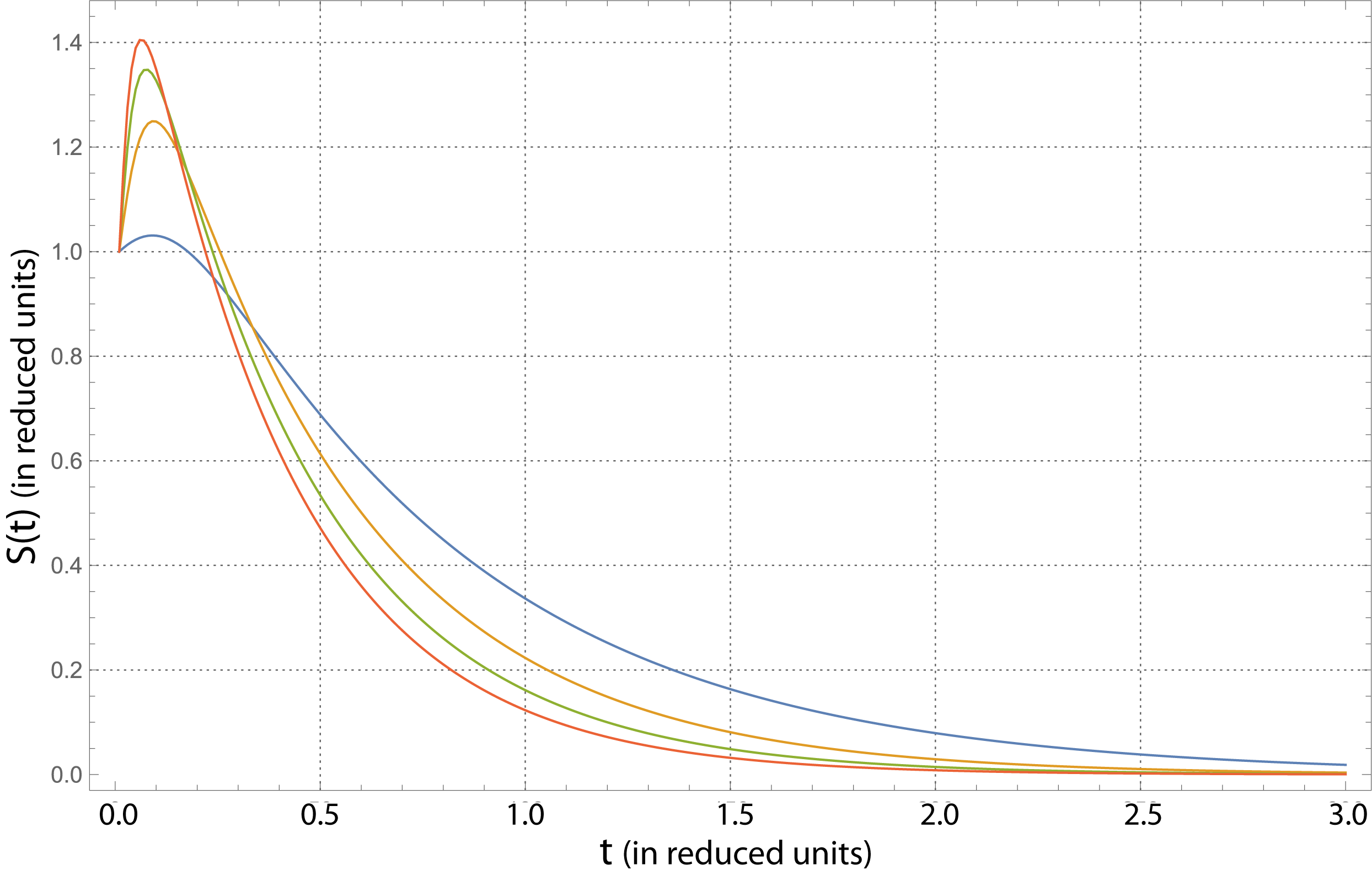} 
\caption{$\epsilon^0=0.5$}
\label{fig:subim3b}
\end{subfigure}
\begin{subfigure}{0.49\textwidth}
\includegraphics[width=\linewidth]{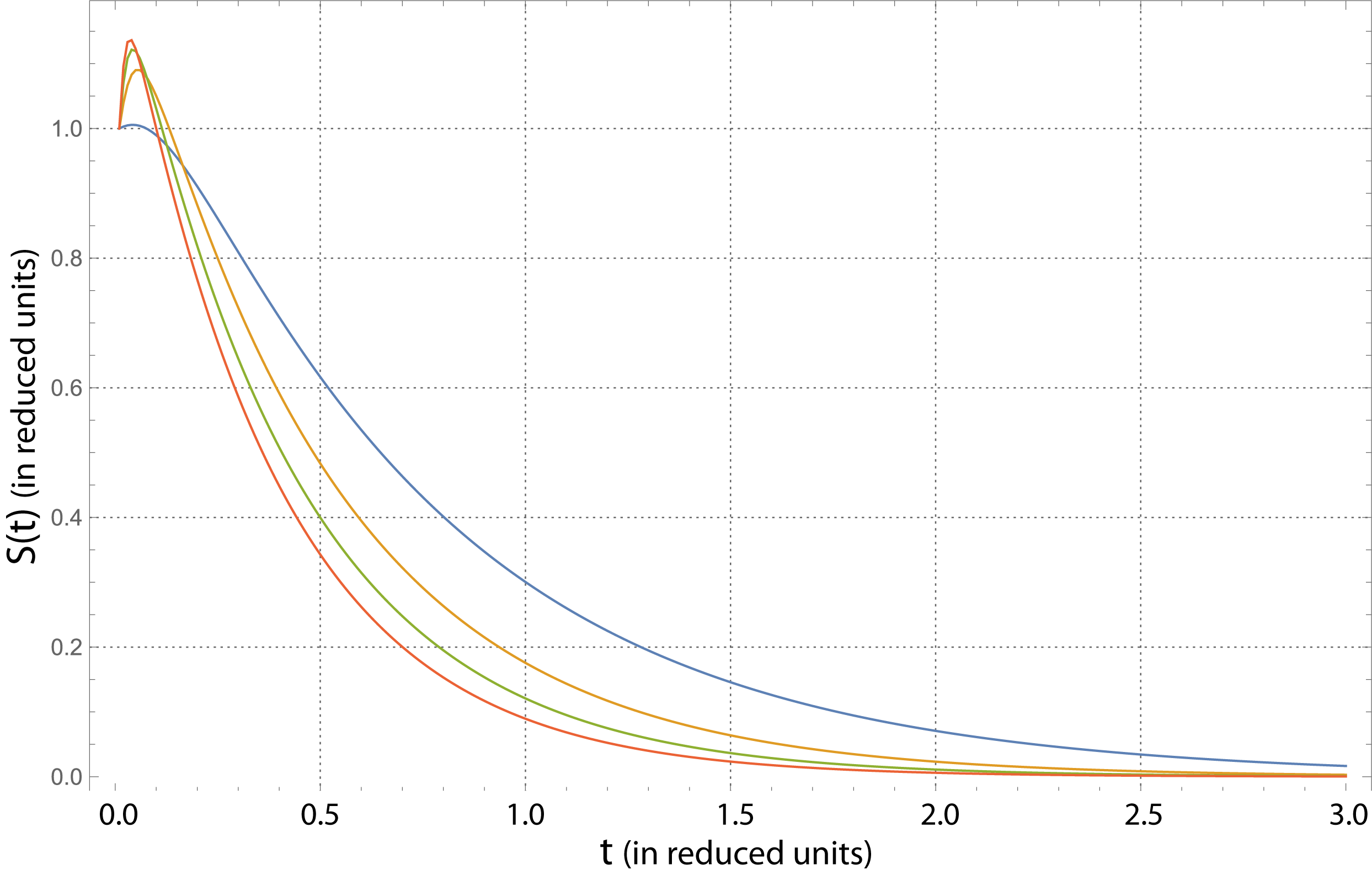}
\caption{$\epsilon^0=0.8$}
\label{fig:subim3c}
\end{subfigure}
\begin{subfigure}{0.49\textwidth}
\includegraphics[width=\linewidth]{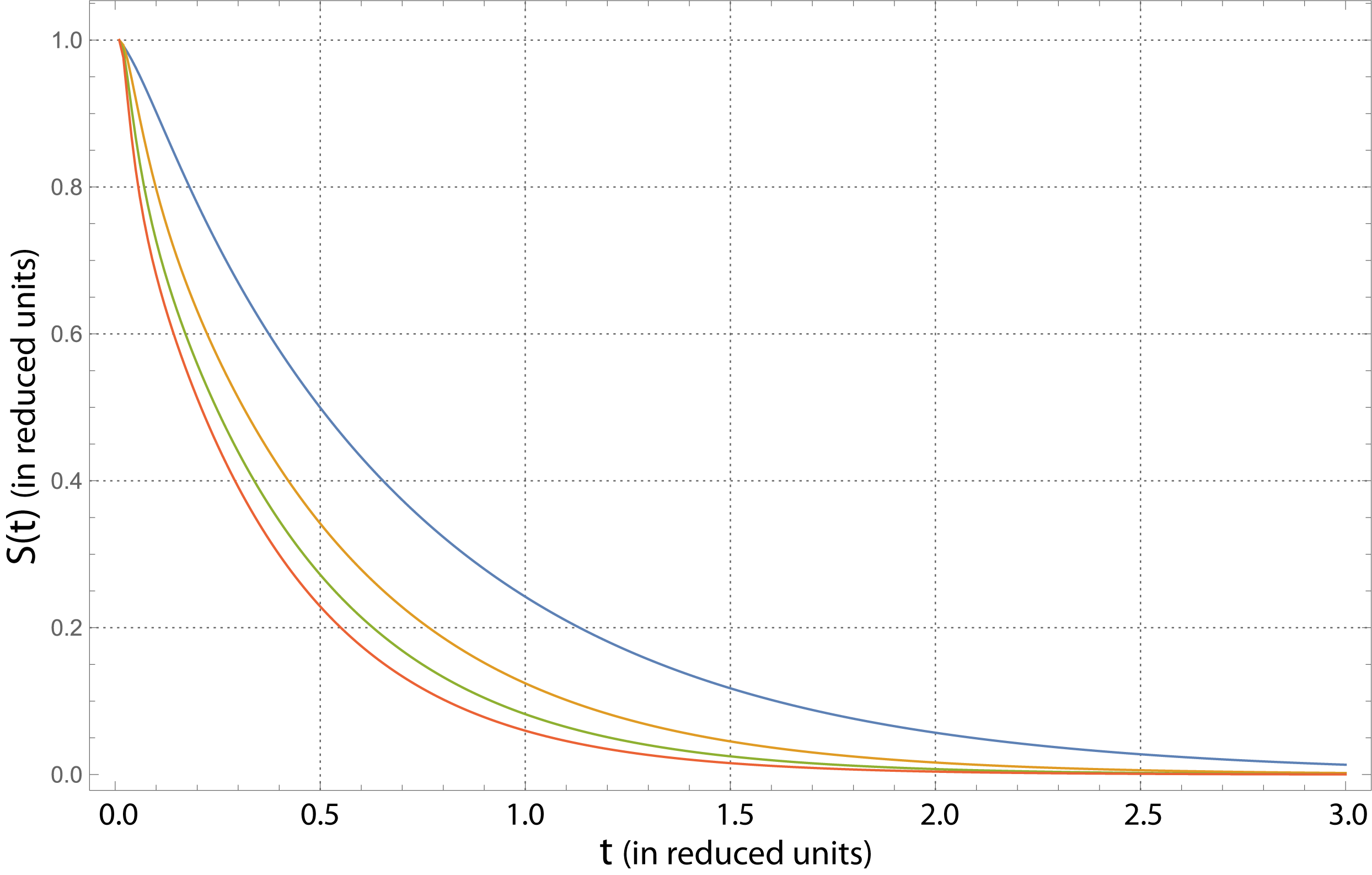}
\caption{$\epsilon^0=1.3$}
\label{fig:subim3d}
\end{subfigure}

\caption{Solvation: Comparison of S(t) vs t for different alpha (1,4,7,10 blue to orange) For various initial wavelet as indicated K 1}
\label{fig:image3}

\end{figure}
\end{widetext}

Figure (\ref{fig:image3}) summarises the behavior of S(t) for various initial preparations $\epsilon^0=0.2,0.5,0.8,1.3$ for $k=1$ as the inhomogenity, $alpha$ is varied. It is seen that the effect is most prominent if the initial condition is prepared near ground state and diminishes with the gap, and beyond a certain finite value of the initial condition becomes monotonic. The effect becomes more prominent with increasing diffusion. In the following section we delve in studying this parametric dependence in detail and construct a phase portrait that indicates the existence of the non-monotonic phase in the parameter space ($\alpha,k$) as well as the relative strength.

\pagebreak

\section{The Phase Portrait}

Figure (\ref{fig:image3}), in previous section portrays dependence of the strength of the non-monotonic behaviour on the strength of the inhomogenity in diffusion ($\alpha$), and the restoring force ($k$), for various initial conditions $\epsilon^0$. In this section we wish to concentrate further on the same and attempt to quantify the behaviour of $S(t)$ for a wide range of these parameters $\alpha,k$.

To begin with we notice the following property that demarcates the non-monotonic phase from the usual monontonic decay. For the parametric combination of $\alpha,k$ where the former exists, it shows an initial growth in $S(t)$ beyond the universal initial value of 1, attains a maximum (to be denoted by $S_{Max}$) and then further relaxes to the equilibrium value. Thus existence of a $S_{Max}$ strictly $>1$ for the initial conditions $\epsilon^0$, denotes a non-monotonic decay, as opposed to the monotonic case, where $S_{Max}$ equals 1 irrespective of all other details.

\begin{widetext}

\begin{figure}[h]
\includegraphics[width=\textwidth]{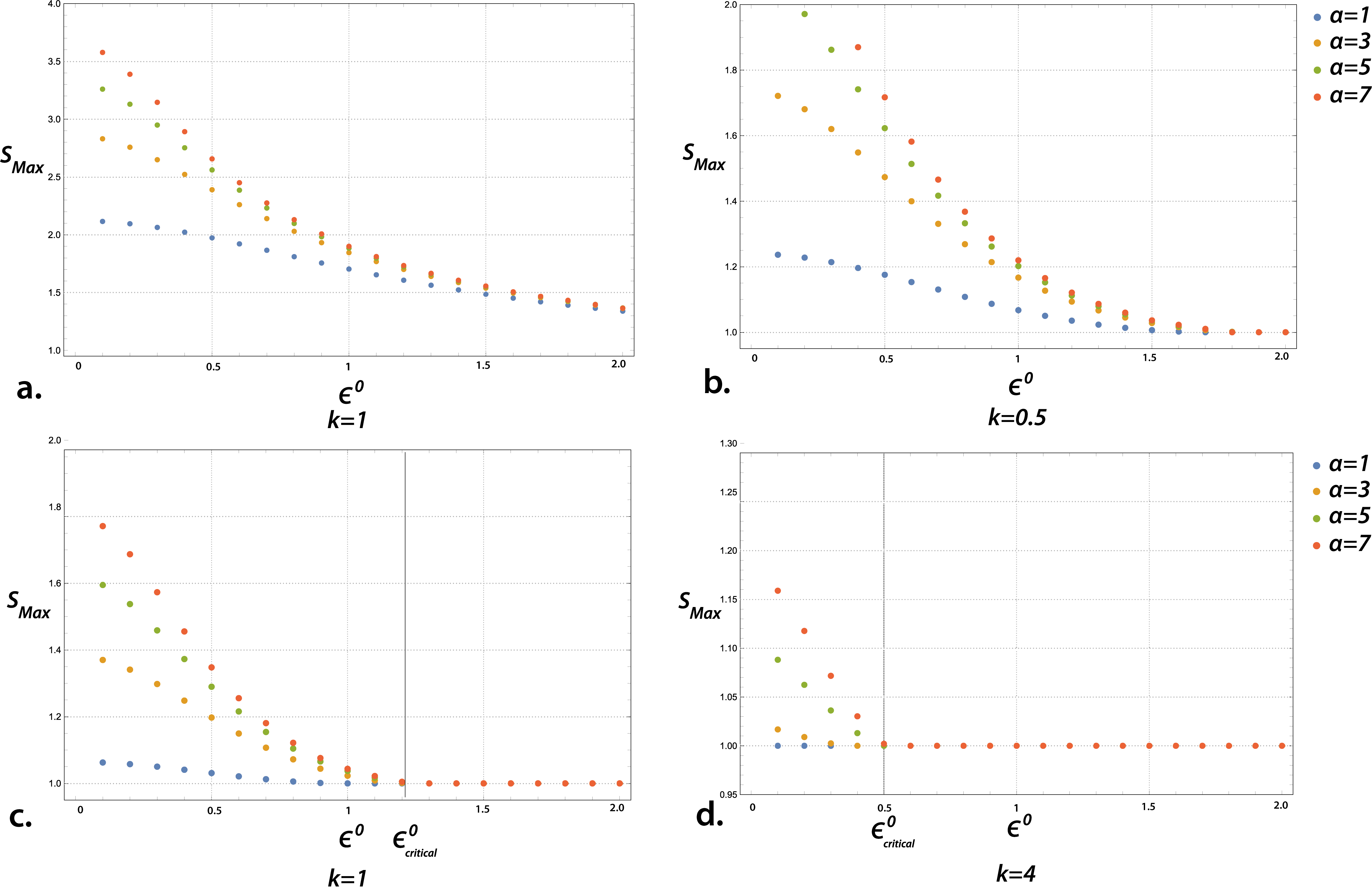} 
\caption{}
\label{fig:image4}
\end{figure}

\end{widetext}

In figure (\ref{fig:image4}), we plot $S_{Max}$ as a function of $\epsilon^0$ in the range [0,2] for various strengths of $\alpha (1,3,5,7)$ and $k (0,0.5,1,4)$. It is seen that the non-monotone nature of the relaxation is strengthened with increase in inhomogeneity and decrease in restoring strength as is denoted by larger values of $S_{Max}$ for given initial condition, as we vary these parameters. The effect is maximised in the free diffusion case $k=0$ as long as we have a finite inhomogeneity $\alpha$. We also notice for a given combination of $\alpha,k$ the range of initial conditions $\epsilon^0$, for which the non-monotonic phase is exhibited has a compact support between 0 and a certain $\epsilon^0_critical$, beyond which the decay is again monotonic. 

\begin{widetext}

\begin{figure}[h]
\includegraphics[width=0.8\textwidth]{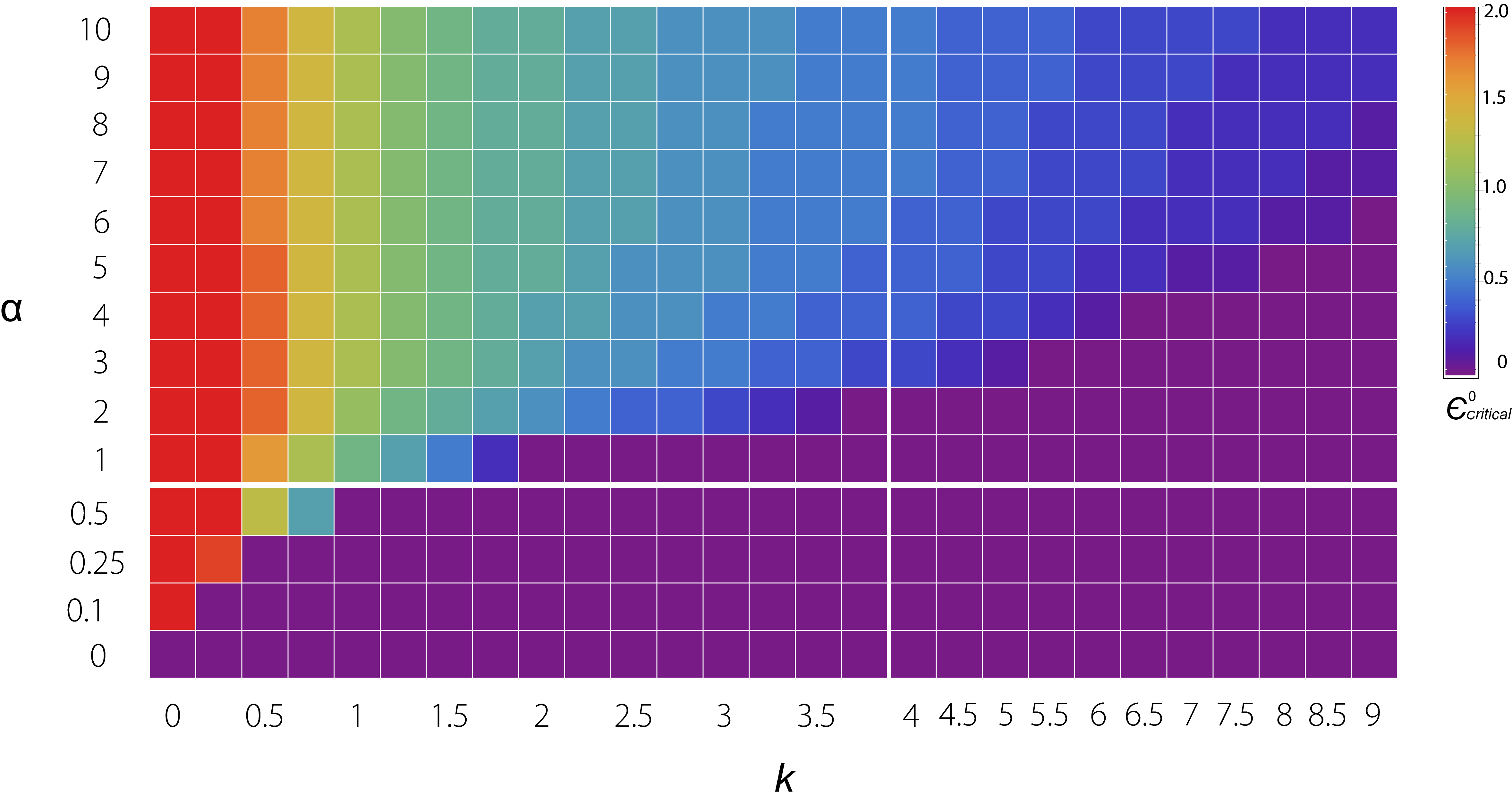} 
\caption{Phase Portrait of $\epsilon_{critical}^0$ against various values of the position dependence ($\alpha$) and the potential strength ($k$), which exhibits the two phases and there phase boundaries. Positive values of $\epsilon_{critical}^0$ (colors above violet) denotes existence of the non-monotonic phase, while a higher value (redder color) indicates its abundance for a given combination of ($\alpha,k$). The effect is stronger for higher diffusion ($\alpha$) and lower potential strength (decreasing $k$), which reasserts its origin in interplay of these two opposing force. $\epsilon_{critical}^0$ is identically zero for the $\alpha = 0$ line, irrespective of potential strength $k$. This corresponds to the space-independent case, where we expect an exponential decay from LRT . The non-monotonic phase is an exclusive feature of introducing non-homogeneous diffusion.}
\label{fig:image5}
\end{figure}

\end{widetext}

This forms our basis of construction of the phase space portrait, Figure (\ref{fig:image5}). For the parameter ranges of ($\alpha,k$), that does not exhibit the non monotonic phase for any initial condition $\epsilon^0$, the $\epsilon^0_critical$ is evaluated to be zero (denoted by violet). This portrait gives detailed account of the existence of the non-monotonic phase and its relative strength as a parametric dependence of inhomogenity and restoring strength. Firstly we observe that $\epsilon^0_critical$ is identically zero in the line $\alpha=0$ which corresponds to the space independent diffusion. Most of the theoretical as well as experimental studies till date has been concentrated in this regime; the reason for the non monotonic decay to be not observed before. The phase portrait also shows the robustness of the existence of this phases across the entire spectrum of $\alpha,k$ and its relative strength has been indicated by a temperature map, as well as the phase boundary with the strictly monotonic phase. These parameters have been indicated in reduced units, and can be converted to realistic values. Figure (\ref{fig:image5}) thus forms a guide for the experiments for the range of $\alpha,k$ to look at for the detection of this predicted phase.

\section{Conclusion}

In summary, we have studied the effects of introducing position-dependent diffusion in non equilibrium dynamics in a bound harmonic potential. The position dependence introduces non-monotonicity in the relaxation of non-equilibrium correlation functions for a wide range of combinations of inhomogenity and restoring strength. With higher inhomogenity in diffusion as well as lower restoring forces this feature is strengthened. This feature is robust in relaxation of all higher odd moments of the energy gap, where the even moments still exhibit an exponential relaxation. We have studied the criticality and constructed a phase portrait which shows that it requires a certain minimum positional inhomogeneity in diffusion for this non-monotonicity to be induced, which is consistent with the exponential relaxation of correlation functions in the position-independent case. Most theoretical as well as experimental works till date have been performed in this limit, causing this phenomena in the context of reactions of condensed phase remain unexplored. The fondest expectation of our reportage of the same, as well as predictions of the parameter range indicating its abundance is to pave way for experiments for for its unearthing.   

\nocite{*}
\bibliography{main}

\end{document}